\documentclass[journal=jacsat,manuscript=article]{achemso}

\usepackage{chemformula} 
\usepackage[T1]{fontenc} 
\usepackage{siunitx}
\usepackage{amsmath}
\usepackage{graphicx}
\usepackage{glossaries}

\newacronym[longplural={two-dimensional hole gases}]{2dhg}{2DHG}{two-dimensional hole gas}
\newacronym{stqubit}{$S$-$T_{0}$-qubit}{singlet-triplet qubit}
\newacronym{dqd}{DQD}{double quantum dot}
\newacronym{psb}{PSB}{Pauli spin blockade}
\newacronym{sqd}{SQD}{single quantum dot}
\newacronym{qw}{QW}{quantum well}

\author{A. Hofmann}
\email{andrea.hofmann@ist.ac.at}
\altaffiliation{Contributed equally}
\author{D. Jirovec}
\email{daniel.jirovec@ist.ac.at}
\altaffiliation{Contributed equally}
\author{M. Borovkov}
\author{I. Prieto}
\affiliation[IST]
{IST Austria, Am Campus 1, 3400 Klosterneuburg}
\author{A. Ballabio}
\author{J. Frigerio}
\author{D. Chrastina}
\author{G. Isella}
\affiliation[PoliMi]
{L-NESS, Department of Physics, Politecnico di Milano, Via Anzani 42, 22100 Como, Italy}
\author{G. Katsaros}
\affiliation[IST]
{IST Austria, Am Campus 1, 3400 Klosterneuburg}

\title[GeQuantumDots]
  {Assessing the potential of Ge/SiGe quantum dots as hosts for singlet-triplet qubits}

\abbreviations{IR,NMR,UV}

\begin{document}


\begin{abstract}
	We study double quantum dots in a Ge/SiGe heterostructure and test their maturity towards singlet-triplet ($S-T_0$) qubits. We demonstrate a large range of tunability, from two single quantum dots to a double quantum dot. We measure Pauli spin blockade and study the anisotropy of the $g$-factor. We use an adjacent quantum dot for sensing charge transitions in the double quantum dot at interest. In conclusion, Ge/SiGe possesses all ingredients necessary for building a singlet-triplet qubit.
\end{abstract}


\section{Introduction}
Germanium has turned out to be a versatile material for the study of physics at the nanoscale. Confinement into lower dimensions has been achieved in the form of epitaxially grown Ge/Si core/shell nanowires~\cite{Lu2005,Roddaro2008,Hu2012,Higginbotham2014,Brauns2016,Brauns2016a} as well as in dome-islands\cite{Katsaros2010,Ares2013} and hut-wires\cite{Zhang2012,Watzinger2016,Watzinger2018,Vukusic2017,Vukusic2018,Xu2019} by controlling the assembly of Ge on Si. Holes are confined in the Ge-rich part of these nanostructure, thus enabling low-dimensional p-type transport. Using appropriate gate-layouts, fully confined valence band states have been used for semiconductor spin qubits with record manipulation time\cite{Watzinger2018}.

In a different approach, building a Ge \gls{qw} by sandwiching it with SiGe barriers has shown to yield high mobilities of $\mu=\SI{1.1e6}{\centi\meter^2/\volt\second}$ at a density $p=\SI{3e11}{\centi\meter^{-2}}$ when  modulation doping it with boron atoms\cite{Dobbie2012}. Even though the mobility is lower in undoped Ge/SiGe heterostructures\cite{Sammak2019} ($\mu=\SI{5e5}{\centi\meter^2/\volt\second}$ at $p=\SI{6e11}{\centi\meter^{-2}}$), the absence of dopands promises a reduction in scattering and charge noise~\cite{Borselli2011}, which will be important for building quantum-dot based qubits~\cite{Loss1998,Koppens2006,Nowack2011,Hanson2007}. Concerning such, recent studies\cite{Lodari2019,Hendrickx2018,Hendrickx2019} show that the large spin--orbit interaction allows for \SI{100}{\mega\hertz} Rabi frequencies in spin-qubits and two-qubit logic, and that the low effective mass ($\approx\SI{0.05}{\electronmass}$ at low density~\cite{Lodari2019}) features sizable orbital energy spacings. Additionally, low hyperfine interaction~\cite{Fischer2008,Testelin2009} inherent for holes in general and heavy-holes~\cite{Fischer2008} hosted in nuclear-spin-free material in particular promises a quiet qubit environment.
 
Here, we take a closer look various other ingredients which will be necessary for further studies of (qubits in) undoped Ge/SiGe heterostructures, in particular towards the realization of \glspl{stqubit}~\cite{Petta2005,Levy2002}. An \gls{stqubit} takes as a basis the singlet and unpolarized triplet states formed by two electron spins each of which sits in one of two coupled \glspl{sqd}. These two states are separated in energy by the exchange interaction, which in turn is a function of the detuning $\epsilon$ and the tunnel coupling $t$. The Rabi oscillation is caused by the energy difference between their symmetric and anti-symmetric superposition. In the absence of hyperfine interaction~\cite{Foletti2009}, this energy difference is dominated by the difference in the $g$-factors in the left and right \gls{sqd}~\cite{Jock2018}. With the large spin--orbit interaction of valence band states in Ge allowing for electrical tunability of the $g$-factor~\cite{Terrazos2018,Morrison2014}, an \gls{stqubit} will be an interesting alternative to the Loss-DiVincenzo spin qubit~\cite{Loss1998}. In this respect, our interest in the present work lies in analyzing the following parameters of Ge quantum dots~\cite{Hardy2019}, which will be important for a \gls{stqubit}: the tunability of the inter-dot tunnel coupling, the presence of \gls{psb}, charge readout with a capacitively coupled sensor and the $g$-factor anisotropy.

\section{Results and discussion}

\subsection{Tunable double quantum dots}
\begin{figure}[t]
	\includegraphics[width=\textwidth]{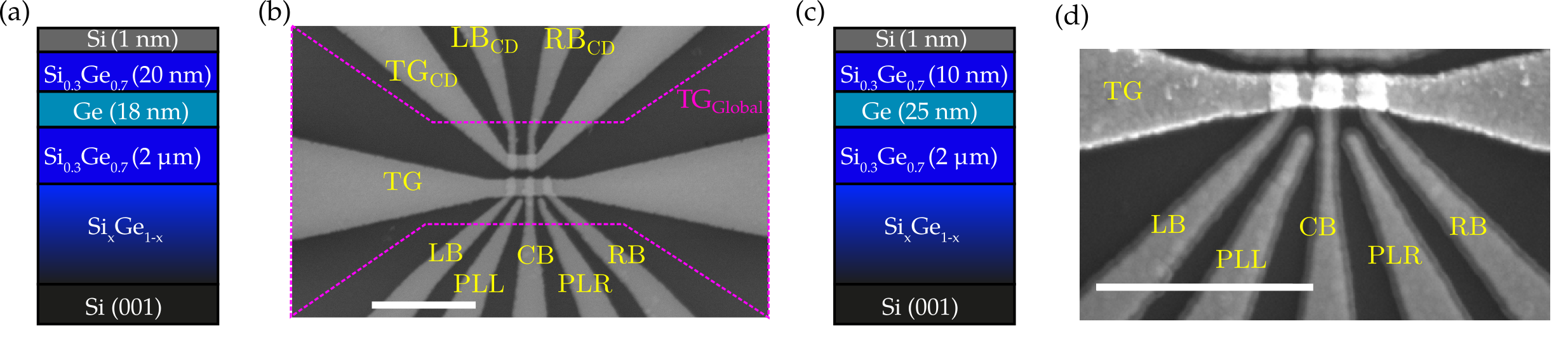}
	\caption{Sample overview. Panels (a-b) describe sample \textbf{A} and panels (c-d) describe sample \textbf{B}. Panels (a,c) display the growth layer stack. Both heterostructures have been grown using low energy plasma enhanced chemical vapor deposition. Standard silicon nanofabrication techniques have been used to fabricate the devices shown in panels (b,d). The scale-bar indicates \SI{1}{\micro\meter}. Gates labeled $TG$ are used as accumulation gates and lie on top of the other gates, with a layer of aluminium oxide in-between. The lower level of gates are used in depletion mode and function as barriers.}
	\label{fig:Layout}
\end{figure}
We present results from two samples originating from two different heterostructures, but similar top-gate layout. The relevant information about sample \textbf{A} and \textbf{B} are summarized in Fig.~\ref{fig:Layout}(a-b) and (c-d), respectively. The active layer in each heterostructure consists of a strained Ge \gls{qw} confined between two strain-relaxed Si$_{0.3}$Ge$_{0.7}$ barrier layers, as shown schematically in Fig.~\ref{fig:Layout}(a,c). This material combination provides an insulating state in equilibrium at cryogenic temperatures but hosts a \gls{2dhg} once the Fermi energy is tuned into the valence band, for example by applying a negative voltage to a metallic top-gate electrode. The \gls{2dhg} of sample \textbf{A} is characterized by a mobility of $\mu=\SI{26000}{\centi\meter^2/\volt\second}$ at a density of $p=\SI{1.1e12}{\centi\meter^{-2}}$ as measured at $\SI{4}{\kelvin}$. The samples are grown using low energy plasma enhanced chemical vapor deposition~\cite{Rossner2004}. In Fig.~\ref{fig:Layout}(b,d), the accumulation gates indicated as $TG$ and $TG_{CD}$ serve to accumulate charges in the \gls{2dhg}, while the finger gates located below the accumulation gates locally screen their electric field and function as barrier gates by depleting the \gls{2dhg} underneath. The additional global top Gate, $TG_{Global}$ in sample \textbf{A} increases the device tunability.

\begin{figure}[t]
	\includegraphics{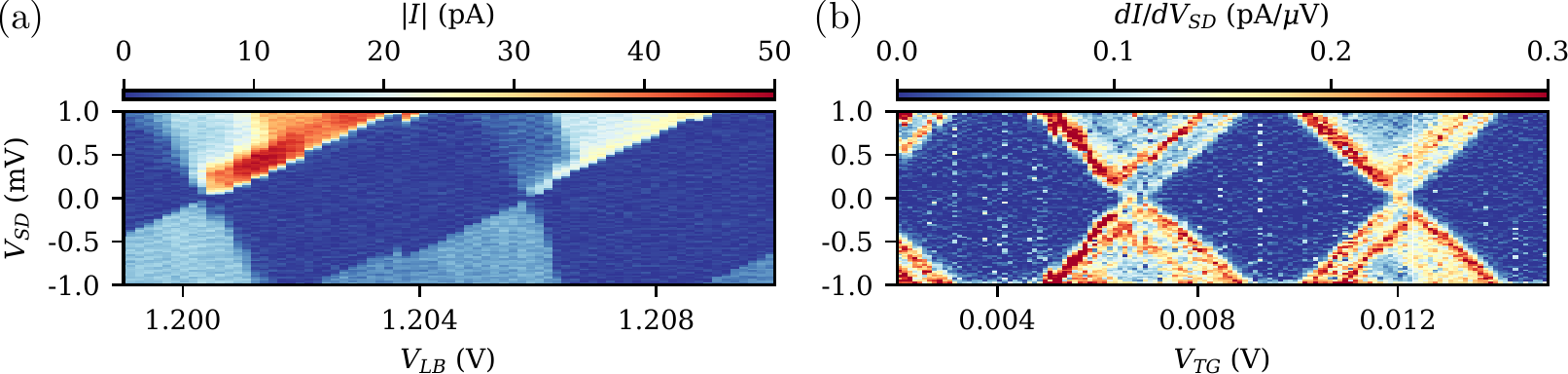}
	\vspace{0ex} \\
	\includegraphics{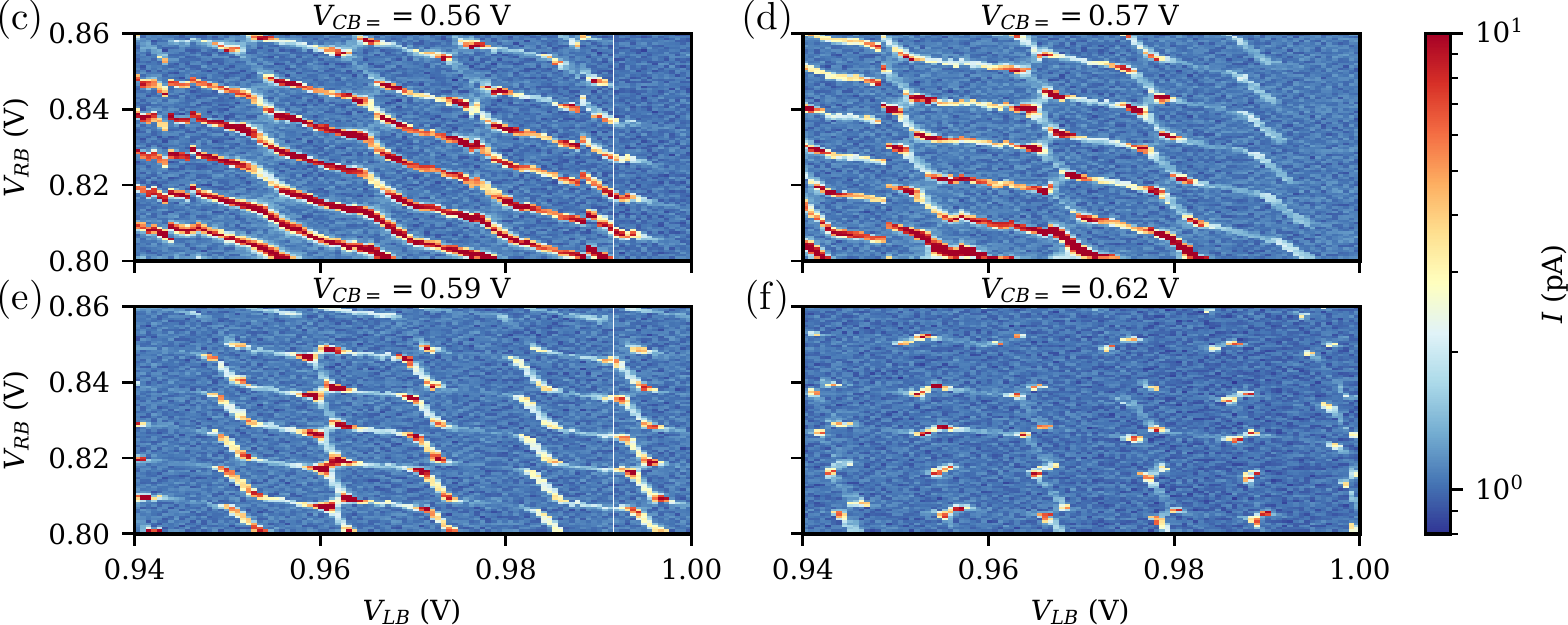}
	\caption{Tunability of Ge/SiGe quantum dots. Panels (a) and (b) each display Coulomb diamonds measured in a \gls{sqd} formed by using, respectively, the left and right set of gates in sample \textbf{B}. Turning on both dots simultaneously, we form a \gls{dqd} shown in panels (c-f). The gate $CB$ controls the tunnel coupling between the two dots, allowing for a tunability ranging from a quantum dot molecule to a \gls{dqd} formed by two isolated \glspl{sqd}.}
	\label{fig:Tunability}
\end{figure}
In order to characterize the sample and show the large phase space available, we use sample \textbf{B} to individually tune two \glspl{sqd} before demonstrating how the center barrier determines the tunnel coupling between these two dots. Fig.~\ref{fig:Tunability}(a) shows Coulomb diamonds of the left \gls{sqd} formed on the left side by leaving the gates $RB$ and $PLR$ unused, i.e. at a constant voltage similar to the one applied to $TG$. On the other hand, Fig.~\ref{fig:Tunability}(b) displays Coulomb diamonds of the right \gls{sqd} formed with gates $LB$ and $PLL$ unused. From the Coulomb diamonds, we determine lever arms $\alpha_L=\SI{0.22}{\electronvolt/\volt}$ and $\alpha_R=\SI{0.26}{\electronvolt/\volt}$, for the left and right \gls{sqd}, respectively, and extract an electron temperature of $T=\SI{50}{\milli\kelvin}$ from the width of a low-bias Coulomb peak measured in the right dot. The charging energy $E_{C}=\SI{1.2}{\milli\volt}$ yields an estimate for the dot radius~\cite{Ihn2009}, $r\approx\frac{e^2}{8\epsilon_0\epsilon E_C}=\SI{117}{\nano\meter}$, which fits the lithographic width of the \gls{sqd}. In the estimate, we used the electron charge, $e$, and permittivities $\varepsilon_0$ in vacuum and $\varepsilon=16.2$ in Ge~\cite{NSM_IOFFE}. In both \glspl{sqd} we observe excited states with typical excitation energies on the order of $\Delta=\SI{150}{\micro\electronvolt}$, which is as well within the regime expected from a simple harmonic oscillator model and using the extracted dot-radius and the light hole effective mass $m^*=\SI{0.09}{\electronmass}$ measured in a similar \gls{2dhg}~\cite{Sammak2019}.

We continue to form a \gls{dqd} by using the left ($LB$) and right ($RB$) barrier gates to tune the coupling to the respective leads and $CB$ to tune the tunnel coupling in-between the two \glspl{sqd}. Starting with $V_{CB}=\SI{0.56}{\volt}$, the large inter-dot tunnel coupling hybridizes the two \glspl{sqd} to a molecule and leads to the charge stability diagram displayed in Fig.~\ref{fig:Tunability}(c). There, the current flows through the device whenever the electrochemical potential of the hybridized quantum dot molecule lies between the source and the drain electrochemical potentials. Upon lowering the inter-dot tunnel coupling by increasing $V_{CB}$, the two dots are being decoupled and the areas where current flows is continuously restricted to the triple points~\cite{Livermore1996}, see Fig.~\ref{fig:Tunability}(d-f). To estimate the change in tunneling conductance, we follow the discussions in Refs.~\citenum{Matveev1996,Livermore1996,Ihn2009}  which conclude that the energies of the individual states can be resolved if the tunneling conductance is smaller than $e^2/h$, where $h$ is the Heisenberg constant. Hence, in the most conductive regime, where we do not resolve individual \gls{sqd} states, the interdot tunneling conductance is larger than $e^2/h$, while it must be much smaller than $e^2/h$ in order to dominate the transport in the least conductive regimes.

\subsection{Pauli spin blockade and charge sensing}
\begin{figure}[th!]
\includegraphics[width=\textwidth]{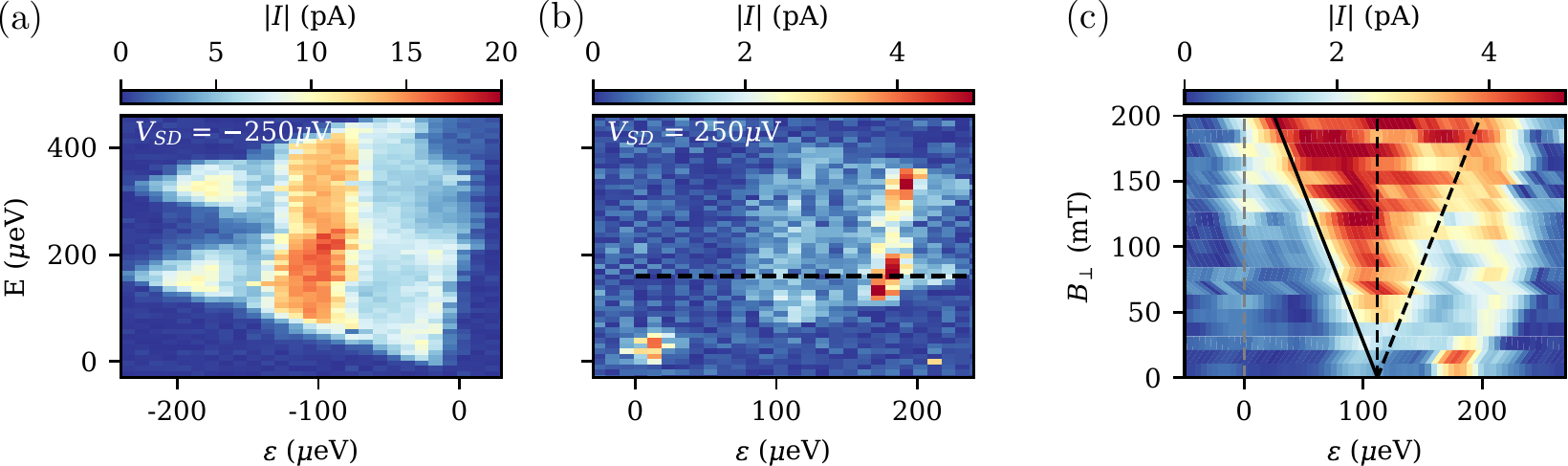}
\vspace{0ex} \\
\includegraphics{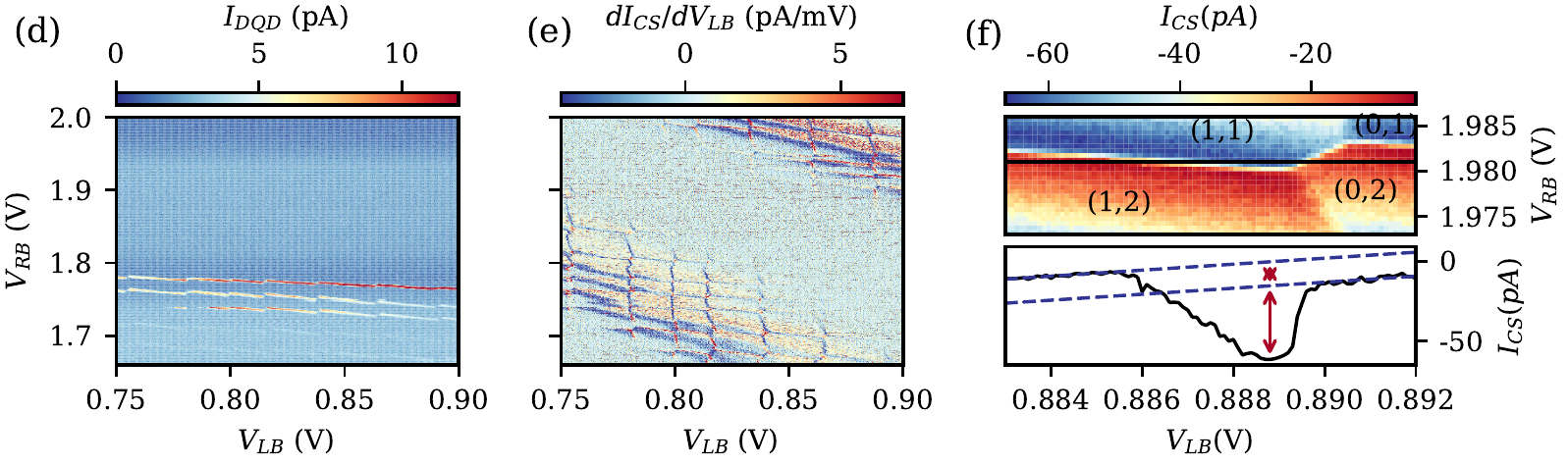}
\caption{Spin blockade and charge readout. Panels (a-c) show the \gls{psb} and panels (d-f) refer to the charge readout scheme. The absolute value of the current through the \gls{dqd} is shown for $V_{SD}=\SI{-250}{\micro\volt}$ in panel (a) and $V_{SD}=\SI{250}{\micro\volt}$ in panel (b), both measured at $B_\perp=\SI{21}{\milli\tesla}$. In (b), the signal is weak along the inter-dot line and enhanced when current starts to flow through the $(0,2)$ triplet state. We repeat these measurements for various values of $B_\perp$, extract traces as indicated by the black dashed line in (b) for each value of the magnetic field and plot them in panel (c). The dashed grey line indicates zero detuning and the black lines follow the position of the $(0,2)$ triplet states, solid for $T_+$ and dashed for $T_0$ and $T_-$. Panels (d) and (e) display the current through the \gls{dqd} and transconductance of charge sensor dot, respectively. Changes in charge states of the \gls{dqd} yield the hexagonal pattern observed in the sensor dot. The Coulomb blockade effect in the sensor dot modulates its sensitivity. A zoom-in of the charge detector current is shown in panel (f,top) together with a line-trace as indicated in black (bottom). Red arrows denote the jumps of differing amplitudes occurring due to electrons tunneling between different charge states of the \gls{dqd}.}
\label{fig:PSB-CD}
\end{figure}
The read-out of a future \gls{stqubit} relies on the \gls{psb}\cite{Hendrickx2019,Ono2002} for spin-to-charge conversion and a charge readout scheme. We will first address the \gls{psb} in Fig.~\ref{fig:PSB-CD}(a-c) before going into detail of charge sensing below. In the experiment, \gls{psb} is characterized by a suppression of current flow through the baseline of the bias triangles in forward bias direction, see Fig.~\ref{fig:PSB-CD}(b), while current is allowed to flow in the reversed direction, as shown in panel (a). Here, forward denotes charge tunnelling through the \gls{dqd} charge states $(0,1)\rightarrow(1,1)\rightarrow(0,2)\rightarrow(0,1)$, where $(N_L,N_R)$ indicates a state with $N_L$ holes in the left and $N_R$ holes in the right dot, where zero is measured from an unknown offset. Then, spin blockade occurs when both, the $(1,1)$ singlet and triplet states are available for transport, while for the $(0,2)$ state, only the singlet ground state is accessible. The blockade is lifted when the detuning equals the $(0,2)$ triplet excitation energy, defined by the orbital level spacing minus the difference in Coulomb energies\cite{Hanson2007}. The resulting singlet-triplet splitting $\Delta_{ST}$ is readily extracted from the onset of current flow in the triangle displayed in Fig.~\ref{fig:PSB-CD}(b), amounting to $\Delta_{ST}=\SI{0.1}{\milli\electronvolt}$ in this case.

Applying a magnetic field perpendicular to the plane of the \gls{2dhg}, $B_\perp>0$, the $(0,2)$ triplet states split with the spin-polarized triplet ($T_+$) state energy decreasing. For small enough magnetic field values where the cyclotron energy $\hbar |e| B_\perp / m^* = 16 B_\perp~ \si{\milli\electronvolt}$ is smaller than the confinement energy (of the order $\hbar^2/m^*(\SI{25}{\nano\meter})^2=\SI{0,1}{\electronvolt}$), we neglect orbital effects in order to approximate $\Delta_{ST}(B_\perp) = \Delta_{ST}(B_\perp=0)-g_\perp\mu_BB_\perp$. This allows us to extract a first estimate of the out-of plane $g$-factor, as we show in Fig.~\ref{fig:PSB-CD}(c). There, we plot in black the line defining $\Delta_{ST}(B_\perp)$ with $g_\perp=7.5$. We will return to discussing $g$-factors in more detail below. Additionally, in Fig.~\ref{fig:PSB-CD}(c), we indicate in black dashed lines how the two remaining triplet states ($T_0$ and $T_-$) evolve with magnetic field.

We now turn to the charge readout mechanism, which relies on the capacitive coupling of the \gls{dqd} under study to a charge-sensitive conductor~\cite{Hanson2007,Field1993}. Here, we use sample \textbf{A} with a \gls{dqd} gate layout similar to sample \textbf{B} but with the addition of a \gls{sqd} in close vicinity to it, see Fig.~\ref{fig:Layout}(b). The current through the additional charge sensor dot, $I_{CS}$, exhibits steps whenever the electronic configuration of the \gls{dqd} is changed. Fig.~\ref{fig:PSB-CD} shows the current $I_{DQD}$ measured through the \gls{dqd} in panel (d) and the numerical $dI_{CS}/dV_{LB}$ through the sensor dot in panel (e), recorded in the same measurement. While the signal $I_{DQD}$ vanishes in the noise background, the transconductance through the charge sensor dot maps the \gls{dqd}'s hexagonal charge stability diagram. The modulation of the signal intensity is caused by Coulomb oscillations in the sensor dot.

A zoom-in of $I_{CS}$ is provided in Fig.~\ref{fig:PSB-CD}(f,top), together with a trace along the line indicated in black (bottom). The step-height of $I_{CS}$ occurring due to charge transitions in the \gls{dqd} depends on the involved charge states. In particular, we clearly distinguish the inter-dot transition $(0,2)-(1,1)$ from the transition $(1,2)-(1,1)$ where an electron is exchanged with a reservoir. Together with the \gls{psb} demonstrated above, this fulfills the requirement of a spin-readout mechanism.

\subsection{$g$-factor anisotropy}
The all-electrical control of a \gls{stqubit} requires different $g$-factors in the two \glspl{sqd} forming the \gls{dqd}. It is therefore interesting to study in detail the $g$-factors, which we do in the following via measuring the splitting of a Kondo peak in an applied magnetic field.

\begin{figure}[t]
	\includegraphics{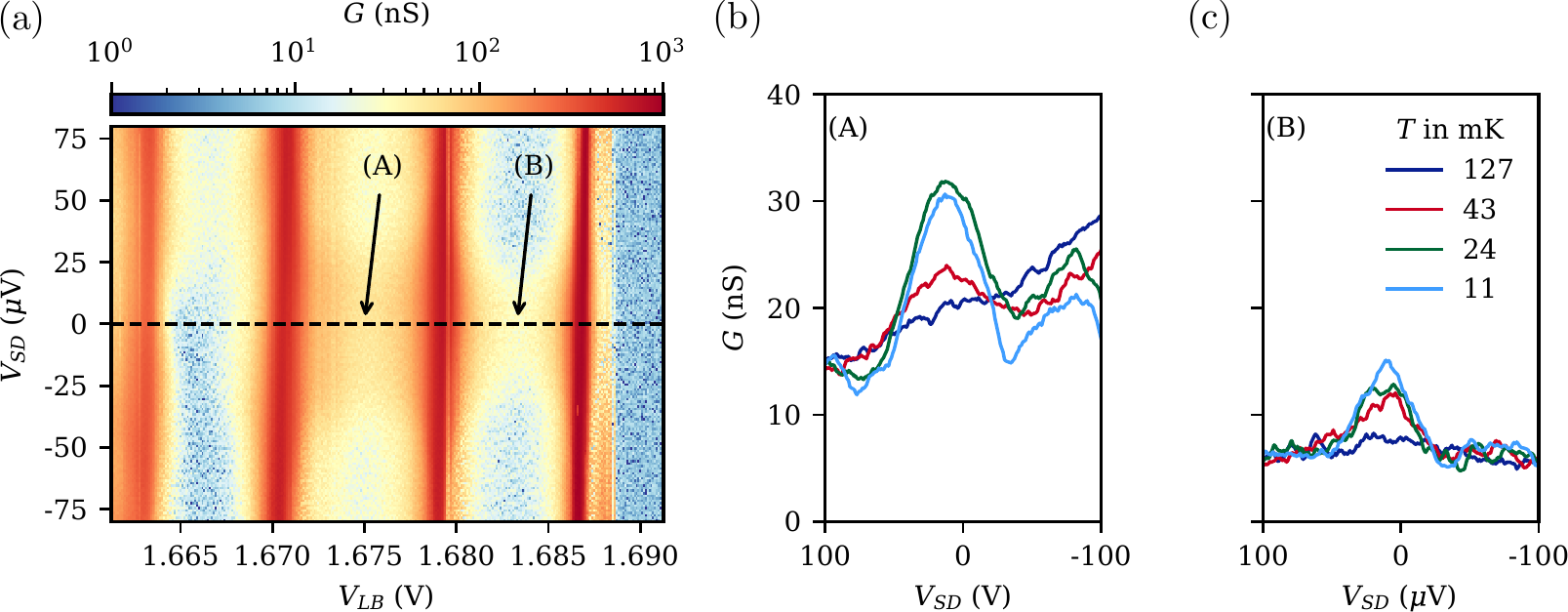}
	\caption{The Kondo effect. The measured (using standard lock-in techniques) conductance as a function of gate and bias voltage, displayed in (a), exhibits peaks and valleys not only as expected due to the Coulomb blockade effect but also in-between Coulomb peaks. The additional peaks are labelled (A) and (B) and are studied as a function of temperature in panels (b) and (c), respecitvely. Their decay with temperature suggests Kondo temperatures smaller than $\SI{127}{\milli\kelvin}$.}
	\label{fig:Kondo}
\end{figure}

In a first step, we tune the left \gls{sqd} of device \textbf{B} into a regime where at least one barrier is transparent enough to observe the Kondo effect \cite{Franceschi2010}. Fig.~\ref{fig:Kondo}(a) displays the conductance $G$ at low bias and varying gate-voltage. It peaks due to Coulomb resonances, and additionally, two regions labeled (A) and (B) exhibit increased conductance at $V_{SD}=0$. Fig~\ref{fig:Kondo}(b) and (c) show line-traces through the conductance peaks (A) and (B), respectively, at fixed gate voltages and varying temperatures. As expected for the Kondo effect, the conductance decreases for increasing temperature and vanishes above the Kondo temperature $T_K$. From the full width at half maximum of the conductance peak (A) at the lowest temperature, we estimate $T_K=\SI{100}{\milli\kelvin}$. The occurrence of Kondo peaks in two consecutive Coulomb-valleys likely results from orbital degeneracies in the quantum dot energy spectrum, similar as observed in Ref.~\citenum{Li2015}.

\begin{figure}[t]
	\includegraphics{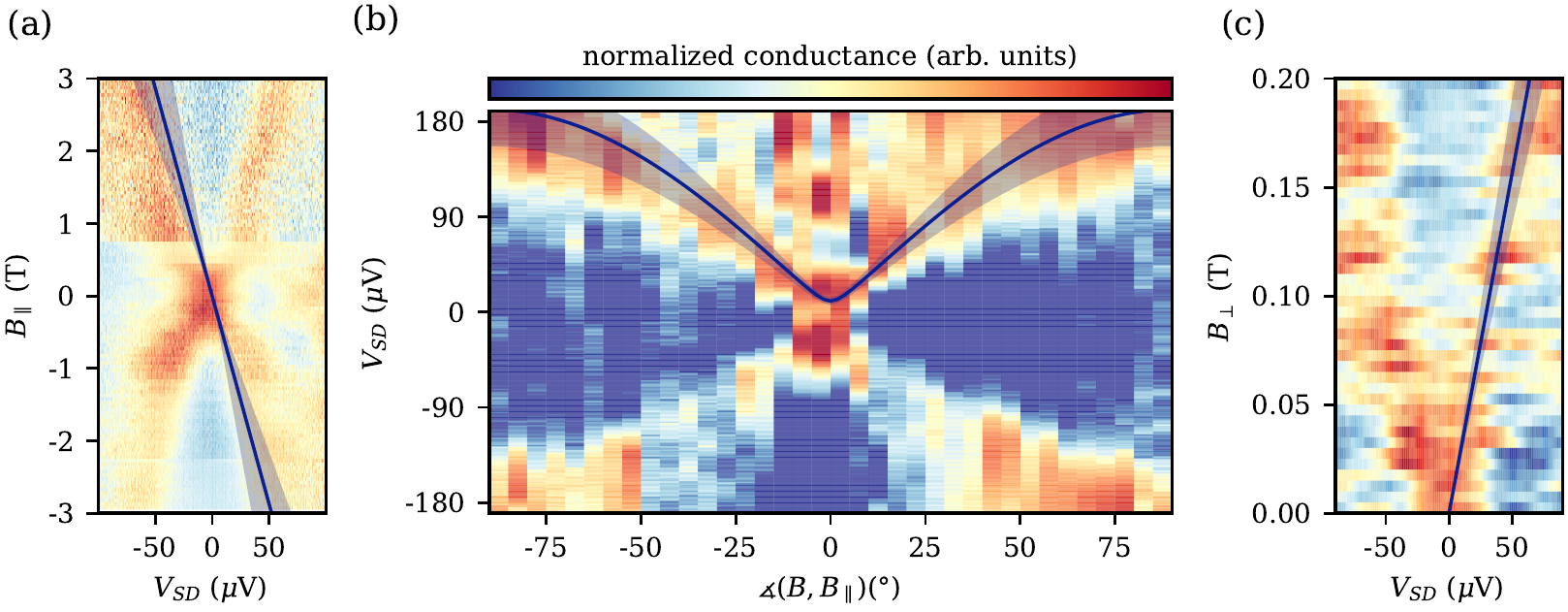}
	\caption{$g$-factor anisotropy. The conductance is shown as a function of in-plane field in (a), as a function of out-of-plane field in (c) and as a function of out-of-plane angle (at $|B|=\SI{0.6}{\tesla}$) in (b). The conductance was measured with standard lock-in techniques and the values measured at different angles have been normalized to a common minimum and maximum. The blue lines display the condition $|e|V_{SD} = g\mu_B B$ for the $g$-factors $g_\parallel=0.3$ and $g_\perp=5.5$. The error bars define the boundaries of the blue shaded regions.}
	\label{fig:Anisotropy}
\end{figure}
We go on to extract the $g$-factors in and out-of plane by using the splitting of peak (A) in an applied magnetic field. With the field parallel to the plane of the \gls{2dhg}, the splitting is characterized by $g_\parallel$, as shown in Fig.~\ref{fig:Anisotropy}(a). Similarly, again neglecting orbital effects, Fig.~\ref{fig:Anisotropy}(c) shows the splitting of the conductance peak due $g_\perp$  and an applied magnetic field perpendicular to the \gls{2dhg}-plane. The $g$-factors extracted from the inflection points of the conductance~\cite{Csonka2008} as measured in panels (a) and (c) are $g_\perp=5.5\pm1.0$ and $g_\parallel=0.3\pm0.1$, respectively, which we plot in blue lines within shaded regions corresponding to the error bars. Finally, we fix the amplitude of the magnetic field at $|B|=\SI{0.6}{\tesla}$ and measure the conductance as a function of the out-of plane angle. The anisotropy of the $g$-factor is visible in the resulting measurement, Fig.~\ref{fig:Anisotropy}(b). Following Refs.~\citenum{Petta2002,Brouwer2000} we write $g(B_\parallel, B_\perp)=\frac{\sqrt{g_\parallel^2 B_\parallel^2 + g_z^2 B_z^2}}{\sqrt{B_\parallel^2+B_\perp^2}}$ for the two-dimensional case at hand. Correspondingly, we indicate in panel (b) using a blue line and a shaded region the behavior of the conductance peaks expected by plugging in the $g_\parallel$ and $g_\perp$ measured as described above. This measurement at constant field underestimates the value for $g_\perp$ probably due to orbital effects starting to play a role at $|B|=\SI{0.6}{\tesla}$.

The measured value of $g_\perp$ is smaller than the theoretically predicted value $g_\perp=21.4$ for pure heavy-hole states~\cite{VanKesteren1990}. Admixture of light hole states due the confinement into zero dimensions or strain generally reduces the $g$-factor~\cite{Ares2013}. However, calculations for hut-wires~\cite{Watzinger2016} show that the confinement alone does not reduce the heavy-hole character by more than \SI{1}{\percent}. More likely, leakage of the confined states into the SiGe barries causes the reduction, as $g=2$ for holes in Si~\cite{Voisin2016}.

The anisotropy of the $g$-factor brings about an interesting tuning knob for a \gls{stqubit}~\cite{Jock2018}. For the rotation along the qubit axis, we rely on a difference in $g$-factors between the left and right \gls{sqd}. It has been shown\cite{Hendrickx2019} that small differences in the percentage range can be achieved. Extrapolating from this, we propose to use the shown anisotropy to chose the absolute value of g and hence, expecting a similar percentage range of tunability, its difference in the two \glspl{sqd}.

\section{Conclusion}
We conclude that Ge/SiGe heterostructures possess all the necessary ingredients for building an all-electrically tunable \gls{stqubit}. A qubit can readily be encoded in the singlet and unpolarized triplet states of a \gls{dqd} built from two individually tunable \glspl{sqd}. Readout of the qubit will occur through charge sensing and the \gls{psb} regime. The rotation along the equator may be tuned by the tunnel-coupling between the dots. For the rotation along the qubit axis we can chose the value of the $g$-factor by correspondingly aligning the magnetic field. Apart from chosing a path towards quantum computing with quantum dots, the large $g$-factors together with the elsewhere demonstrated coupling to superconductors~\cite{Vigneau2019,Hendrickx2019a} and long ballistic one-dimensional channels\cite{Mizokuchi2018} make Ge/SiGe an interesting material for studying topologically protected states~\cite{Lutchyn2010,Oreg2010}.

\section{Fabrication of the devices}
\label{sec:fab}
A \SI{90}{\second} long etching process in an SF$_6$-O$_2$-CHF$_3$ reactive ion plasma leaves a mesa on the heterostructure. The etch depth amounts to $\approx\SI{70}{\nano\meter}$. Electric contacts to the Ge consist of a \SI{60}{\nano\meter} thick layer of Pt deposited after removing the native oxide with an in-situ Ar milling process ($2\times2$\si{\minute} at \SI{10}{\milli\ampere} and \SI{300}{\volt}) in the same machine and without breaking the vacuum. These contacts are covered with $200$~cycles ($\approx\SI{20}{\nano\meter}$) of thermal aluminium oxide grown in an ALD at \SI{300}{\celsius}. The depletion gates consist of \SI{3}{\nano\meter} Ti and \SI{92}{\nano\meter} Pd. A second layer of the same oxide isolates the depletion gates from the accumulation gates, which consist of \SI{3}{\nano\meter} Ti and \SI{102}{\nano\meter} Pd. The patterns for all layers are defined using electron-beam lithography at \SI{100}{\kilo\electronvolt}.

\begin{acknowledgement}
We thank Matthias Brauns for helpful discussions and careful proofreading of the manuscript.
This project has received funding from the European Union's Horizon 2020 research and innovation program under the Marie Sklodowska-Curie grant agreement No 844511 and from the FWF project P30207. The research was supported by the Scientific Service Units of IST Austria through resources provided by the MIBA machine shop and the nanofabrication facility.
\end{acknowledgement}

\section{Note}
While writing this manuscript, we became aware of an experiment~\cite{Lawrie2019} where a Ge quantum dot based on a similar heterostructure as the one used in this study shows charge sensing of a quantum dot filled to the last hole.

\bibliography{./../library}

\providecommand{\latin}[1]{#1}
\makeatletter
\providecommand{\doi}
  {\begingroup\let\do\@makeother\dospecials
  \catcode`\{=1 \catcode`\}=2 \doi@aux}
\providecommand{\doi@aux}[1]{\endgroup\texttt{#1}}
\makeatother
\providecommand*\mcitethebibliography{\thebibliography}
\csname @ifundefined\endcsname{endmcitethebibliography}
  {\let\endmcitethebibliography\endthebibliography}{}
\begin{mcitethebibliography}{54}
\providecommand*\natexlab[1]{#1}
\providecommand*\mciteSetBstSublistMode[1]{}
\providecommand*\mciteSetBstMaxWidthForm[2]{}
\providecommand*\mciteBstWouldAddEndPuncttrue
  {\def\EndOfBibitem{\unskip.}}
\providecommand*\mciteBstWouldAddEndPunctfalse
  {\let\EndOfBibitem\relax}
\providecommand*\mciteSetBstMidEndSepPunct[3]{}
\providecommand*\mciteSetBstSublistLabelBeginEnd[3]{}
\providecommand*\EndOfBibitem{}
\mciteSetBstSublistMode{f}
\mciteSetBstMaxWidthForm{subitem}{(\alph{mcitesubitemcount})}
\mciteSetBstSublistLabelBeginEnd
  {\mcitemaxwidthsubitemform\space}
  {\relax}
  {\relax}

\bibitem[Lu \latin{et~al.}(2005)Lu, Xiang, Timko, Wu, and Lieber]{Lu2005}
Lu,~W.; Xiang,~J.; Timko,~B.~P.; Wu,~Y.; Lieber,~C.~M. {One-dimensional hole
  gas in germanium/silicon nanowire heterostructures}. \emph{Proceedings of the
  National Academy of Sciences} \textbf{2005}, \emph{102}, 10046--10051\relax
\mciteBstWouldAddEndPuncttrue
\mciteSetBstMidEndSepPunct{\mcitedefaultmidpunct}
{\mcitedefaultendpunct}{\mcitedefaultseppunct}\relax
\EndOfBibitem
\bibitem[Roddaro \latin{et~al.}(2008)Roddaro, Fuhrer, Brusheim, Fasth, Xu,
  Samuelson, Xiang, and Lieber]{Roddaro2008}
Roddaro,~S.; Fuhrer,~A.; Brusheim,~P.; Fasth,~C.; Xu,~H.~Q.; Samuelson,~L.;
  Xiang,~J.; Lieber,~C.~M. {Spin States of Holes in Ge / Si Nanowire Quantum
  Dots}. \emph{Physical Review Letters} \textbf{2008}, \emph{101}, 186802\relax
\mciteBstWouldAddEndPuncttrue
\mciteSetBstMidEndSepPunct{\mcitedefaultmidpunct}
{\mcitedefaultendpunct}{\mcitedefaultseppunct}\relax
\EndOfBibitem
\bibitem[Hu \latin{et~al.}(2012)Hu, Kuemmeth, Lieber, and Marcus]{Hu2012}
Hu,~Y.; Kuemmeth,~F.; Lieber,~C.~M.; Marcus,~C.~M. {Hole spin relaxation in
  Ge-Si core-shell nanowire qubits}. \emph{Nature Nanotechnology}
  \textbf{2012}, \emph{7}, 47--50\relax
\mciteBstWouldAddEndPuncttrue
\mciteSetBstMidEndSepPunct{\mcitedefaultmidpunct}
{\mcitedefaultendpunct}{\mcitedefaultseppunct}\relax
\EndOfBibitem
\bibitem[Higginbotham \latin{et~al.}(2014)Higginbotham, Kuemmeth, Larsen,
  Fitzpatrick, Yao, Yan, Lieber, and Marcus]{Higginbotham2014}
Higginbotham,~A.; Kuemmeth,~F.; Larsen,~T.; Fitzpatrick,~M.; Yao,~J.; Yan,~H.;
  Lieber,~C.; Marcus,~C. {Antilocalization of Coulomb Blockade in a Ge/Si
  Nanowire}. \emph{Physical Review Letters} \textbf{2014}, \emph{112},
  216806\relax
\mciteBstWouldAddEndPuncttrue
\mciteSetBstMidEndSepPunct{\mcitedefaultmidpunct}
{\mcitedefaultendpunct}{\mcitedefaultseppunct}\relax
\EndOfBibitem
\bibitem[Brauns \latin{et~al.}(2016)Brauns, Ridderbos, Li, Bakkers, and
  Zwanenburg]{Brauns2016}
Brauns,~M.; Ridderbos,~J.; Li,~A.; Bakkers,~E. P. A.~M.; Zwanenburg,~F.~A.
  {Electric-field dependent g -factor anisotropy in Ge-Si core-shell nanowire
  quantum dots}. \emph{Physical Review B} \textbf{2016}, \emph{93},
  121408\relax
\mciteBstWouldAddEndPuncttrue
\mciteSetBstMidEndSepPunct{\mcitedefaultmidpunct}
{\mcitedefaultendpunct}{\mcitedefaultseppunct}\relax
\EndOfBibitem
\bibitem[Brauns \latin{et~al.}(2016)Brauns, Ridderbos, Li, Bakkers, van~der
  Wiel, and Zwanenburg]{Brauns2016a}
Brauns,~M.; Ridderbos,~J.; Li,~A.; Bakkers,~E. P. A.~M.; van~der Wiel,~W.~G.;
  Zwanenburg,~F.~A. {Anisotropic Pauli spin blockade in hole quantum dots}.
  \emph{Physical Review B} \textbf{2016}, \emph{94}, 041411\relax
\mciteBstWouldAddEndPuncttrue
\mciteSetBstMidEndSepPunct{\mcitedefaultmidpunct}
{\mcitedefaultendpunct}{\mcitedefaultseppunct}\relax
\EndOfBibitem
\bibitem[Katsaros \latin{et~al.}(2010)Katsaros, Spathis, Stoffel, Fournel,
  Mongillo, Bouchiat, Lefloch, Rastelli, Schmidt, and {De
  Franceschi}]{Katsaros2010}
Katsaros,~G.; Spathis,~P.; Stoffel,~M.; Fournel,~F.; Mongillo,~M.;
  Bouchiat,~V.; Lefloch,~F.; Rastelli,~A.; Schmidt,~O.~G.; {De Franceschi},~S.
  {Hybrid superconductor-semiconductor devices made from self-assembled SiGe
  nanocrystals on silicon}. \emph{Nature Nanotechnology} \textbf{2010},
  \emph{5}, 458--464\relax
\mciteBstWouldAddEndPuncttrue
\mciteSetBstMidEndSepPunct{\mcitedefaultmidpunct}
{\mcitedefaultendpunct}{\mcitedefaultseppunct}\relax
\EndOfBibitem
\bibitem[Ares \latin{et~al.}(2013)Ares, Golovach, Katsaros, Stoffel, Fournel,
  Glazman, Schmidt, and {De Franceschi}]{Ares2013}
Ares,~N.; Golovach,~V.~N.; Katsaros,~G.; Stoffel,~M.; Fournel,~F.;
  Glazman,~L.~I.; Schmidt,~O.~G.; {De Franceschi},~S. {Nature of Tunable Hole g
  Factors in Quantum Dots}. \emph{Physical Review Letters} \textbf{2013},
  \emph{110}, 046602\relax
\mciteBstWouldAddEndPuncttrue
\mciteSetBstMidEndSepPunct{\mcitedefaultmidpunct}
{\mcitedefaultendpunct}{\mcitedefaultseppunct}\relax
\EndOfBibitem
\bibitem[Zhang \latin{et~al.}(2012)Zhang, Katsaros, Montalenti, Scopece,
  Rezaev, Mickel, Rellinghaus, Miglio, {De Franceschi}, Rastelli, and
  Schmidt]{Zhang2012}
Zhang,~J.~J.; Katsaros,~G.; Montalenti,~F.; Scopece,~D.; Rezaev,~R.~O.;
  Mickel,~C.; Rellinghaus,~B.; Miglio,~L.; {De Franceschi},~S.; Rastelli,~A.;
  Schmidt,~O.~G. {Monolithic Growth of Ultrathin Ge Nanowires on Si(001)}.
  \emph{Physical Review Letters} \textbf{2012}, \emph{109}, 085502\relax
\mciteBstWouldAddEndPuncttrue
\mciteSetBstMidEndSepPunct{\mcitedefaultmidpunct}
{\mcitedefaultendpunct}{\mcitedefaultseppunct}\relax
\EndOfBibitem
\bibitem[Watzinger \latin{et~al.}(2016)Watzinger, Kloeffel,
  Vuku{\v{s}}i{\'{c}}, Rossell, Sessi, Kuku{\v{c}}ka, Kirchschlager, Lausecker,
  Truhlar, Glaser, Rastelli, Fuhrer, Loss, and Katsaros]{Watzinger2016}
Watzinger,~H.; Kloeffel,~C.; Vuku{\v{s}}i{\'{c}},~L.; Rossell,~M.~D.;
  Sessi,~V.; Kuku{\v{c}}ka,~J.; Kirchschlager,~R.; Lausecker,~E.; Truhlar,~A.;
  Glaser,~M.; Rastelli,~A.; Fuhrer,~A.; Loss,~D.; Katsaros,~G. {Heavy-Hole
  States in Germanium Hut Wires}. \emph{Nano Letters} \textbf{2016}, \emph{16},
  6879--6885\relax
\mciteBstWouldAddEndPuncttrue
\mciteSetBstMidEndSepPunct{\mcitedefaultmidpunct}
{\mcitedefaultendpunct}{\mcitedefaultseppunct}\relax
\EndOfBibitem
\bibitem[Watzinger \latin{et~al.}(2018)Watzinger, Kuku{\v{c}}ka,
  Vuku{\v{s}}i{\'{c}}, Gao, Wang, Sch{\"{a}}ffler, Zhang, and
  Katsaros]{Watzinger2018}
Watzinger,~H.; Kuku{\v{c}}ka,~J.; Vuku{\v{s}}i{\'{c}},~L.; Gao,~F.; Wang,~T.;
  Sch{\"{a}}ffler,~F.; Zhang,~J.-J.; Katsaros,~G. {A germanium hole spin
  qubit}. \emph{Nature Communications} \textbf{2018}, \emph{9}, 3902\relax
\mciteBstWouldAddEndPuncttrue
\mciteSetBstMidEndSepPunct{\mcitedefaultmidpunct}
{\mcitedefaultendpunct}{\mcitedefaultseppunct}\relax
\EndOfBibitem
\bibitem[Vuku{\v{s}}i{\'{c}} \latin{et~al.}(2017)Vuku{\v{s}}i{\'{c}},
  Kuku{\v{c}}ka, Watzinger, and Katsaros]{Vukusic2017}
Vuku{\v{s}}i{\'{c}},~L.; Kuku{\v{c}}ka,~J.; Watzinger,~H.; Katsaros,~G. {Fast
  Hole Tunneling Times in Germanium Hut Wires Probed by Single-Shot
  Reflectometry}. \emph{Nano Letters} \textbf{2017}, \emph{17},
  5706--5710\relax
\mciteBstWouldAddEndPuncttrue
\mciteSetBstMidEndSepPunct{\mcitedefaultmidpunct}
{\mcitedefaultendpunct}{\mcitedefaultseppunct}\relax
\EndOfBibitem
\bibitem[Vuku{\v{s}}i{\'{c}} \latin{et~al.}(2018)Vuku{\v{s}}i{\'{c}},
  Kuku{\v{c}}ka, Watzinger, Milem, Sch{\"{a}}ffler, and Katsaros]{Vukusic2018}
Vuku{\v{s}}i{\'{c}},~L.; Kuku{\v{c}}ka,~J.; Watzinger,~H.; Milem,~J.~M.;
  Sch{\"{a}}ffler,~F.; Katsaros,~G. {Single-Shot Readout of Hole Spins in Ge}.
  \emph{Nano Letters} \textbf{2018}, \emph{18}, 7141--7145\relax
\mciteBstWouldAddEndPuncttrue
\mciteSetBstMidEndSepPunct{\mcitedefaultmidpunct}
{\mcitedefaultendpunct}{\mcitedefaultseppunct}\relax
\EndOfBibitem
\bibitem[Xu \latin{et~al.}(2019)Xu, Li, Gao, Li, Liu, Wang, Cao, Xiao, Wang,
  Zhang, Guo, and Guo]{Xu2019}
Xu,~G.; Li,~Y.; Gao,~F.; Li,~H.-O.; Liu,~H.; Wang,~K.; Cao,~G.; Xiao,~M.;
  Wang,~T.; Zhang,~J.-J.; Guo,~G.-C.; Guo,~G.-P. {Dipole coupling of a tunable
  hole double quantum dot in germanium hut wire to a microwave resonator}.
  \emph{arXiv:1905.01586} \textbf{2019}, \relax
\mciteBstWouldAddEndPunctfalse
\mciteSetBstMidEndSepPunct{\mcitedefaultmidpunct}
{}{\mcitedefaultseppunct}\relax
\EndOfBibitem
\bibitem[Dobbie \latin{et~al.}(2012)Dobbie, Myronov, Morris, Hassan, Prest,
  Shah, Parker, Whall, and Leadley]{Dobbie2012}
Dobbie,~A.; Myronov,~M.; Morris,~R. J.~H.; Hassan,~A. H.~A.; Prest,~M.~J.;
  Shah,~V.~A.; Parker,~E. H.~C.; Whall,~T.~E.; Leadley,~D.~R. {Ultra-high hole
  mobility exceeding one million in a strained germanium quantum well}.
  \emph{Applied Physics Letters} \textbf{2012}, \emph{101}, 172108\relax
\mciteBstWouldAddEndPuncttrue
\mciteSetBstMidEndSepPunct{\mcitedefaultmidpunct}
{\mcitedefaultendpunct}{\mcitedefaultseppunct}\relax
\EndOfBibitem
\bibitem[Sammak \latin{et~al.}(2019)Sammak, Sabbagh, Hendrickx, Lodari,
  {Paquelet Wuetz}, Tosato, Yeoh, Bollani, Virgilio, Schubert, Zaumseil,
  Capellini, Veldhorst, and Scappucci]{Sammak2019}
Sammak,~A.; Sabbagh,~D.; Hendrickx,~N.~W.; Lodari,~M.; {Paquelet Wuetz},~B.;
  Tosato,~A.; Yeoh,~L.; Bollani,~M.; Virgilio,~M.; Schubert,~M.~A.;
  Zaumseil,~P.; Capellini,~G.; Veldhorst,~M.; Scappucci,~G. {Shallow and
  Undoped Germanium Quantum Wells: A Playground for Spin and Hybrid Quantum
  Technology}. \emph{Advanced Functional Materials} \textbf{2019}, \emph{29},
  1807613\relax
\mciteBstWouldAddEndPuncttrue
\mciteSetBstMidEndSepPunct{\mcitedefaultmidpunct}
{\mcitedefaultendpunct}{\mcitedefaultseppunct}\relax
\EndOfBibitem
\bibitem[Borselli \latin{et~al.}(2011)Borselli, Eng, Croke, Maune, Huang, Ross,
  Kiselev, Deelman, Alvarado-Rodriguez, Schmitz, Sokolich, Holabird, Hazard,
  Gyure, and Hunter]{Borselli2011}
Borselli,~M.~G.; Eng,~K.; Croke,~E.~T.; Maune,~B.~M.; Huang,~B.; Ross,~R.~S.;
  Kiselev,~A.~A.; Deelman,~P.~W.; Alvarado-Rodriguez,~I.; Schmitz,~A.~E.;
  Sokolich,~M.; Holabird,~K.~S.; Hazard,~T.~M.; Gyure,~M.~F.; Hunter,~A.~T.
  {Pauli spin blockade in undoped Si/SiGe two-electron double quantum dots}.
  \emph{Applied Physics Letters} \textbf{2011}, \emph{99}, 063109\relax
\mciteBstWouldAddEndPuncttrue
\mciteSetBstMidEndSepPunct{\mcitedefaultmidpunct}
{\mcitedefaultendpunct}{\mcitedefaultseppunct}\relax
\EndOfBibitem
\bibitem[Loss and DiVincenzo(1998)Loss, and DiVincenzo]{Loss1998}
Loss,~D.; DiVincenzo,~D.~P. {Quantum computation with quantum dots}.
  \emph{Physical Review A} \textbf{1998}, \emph{57}, 120--126\relax
\mciteBstWouldAddEndPuncttrue
\mciteSetBstMidEndSepPunct{\mcitedefaultmidpunct}
{\mcitedefaultendpunct}{\mcitedefaultseppunct}\relax
\EndOfBibitem
\bibitem[Koppens \latin{et~al.}(2006)Koppens, Buizert, Tielrooij, Vink, Nowack,
  Meunier, Kouwenhoven, and Vandersypen]{Koppens2006}
Koppens,~F. H.~L.; Buizert,~C.; Tielrooij,~K.~J.; Vink,~I.~T.; Nowack,~K.~C.;
  Meunier,~T.; Kouwenhoven,~L.~P.; Vandersypen,~L. M.~K. {Driven coherent
  oscillations of a single electron spin in a quantum dot}. \emph{Nature}
  \textbf{2006}, \emph{442}, 766--771\relax
\mciteBstWouldAddEndPuncttrue
\mciteSetBstMidEndSepPunct{\mcitedefaultmidpunct}
{\mcitedefaultendpunct}{\mcitedefaultseppunct}\relax
\EndOfBibitem
\bibitem[Nowack \latin{et~al.}(2011)Nowack, Shafiei, Laforest, Prawiroatmodjo,
  Schreiber, Reichl, Wegscheider, and Vandersypen]{Nowack2011}
Nowack,~K.~C.; Shafiei,~M.; Laforest,~M.; Prawiroatmodjo,~G. E. D.~K.;
  Schreiber,~L.~R.; Reichl,~C.; Wegscheider,~W.; Vandersypen,~L. M.~K.
  {Single-Shot Correlations and Two-Qubit Gate of Solid-State Spins}.
  \emph{Science} \textbf{2011}, \emph{333}, 1269--1272\relax
\mciteBstWouldAddEndPuncttrue
\mciteSetBstMidEndSepPunct{\mcitedefaultmidpunct}
{\mcitedefaultendpunct}{\mcitedefaultseppunct}\relax
\EndOfBibitem
\bibitem[Hanson \latin{et~al.}(2007)Hanson, Kouwenhoven, Petta, Tarucha, and
  Vandersypen]{Hanson2007}
Hanson,~R.; Kouwenhoven,~L.~P.; Petta,~J.~R.; Tarucha,~S.; Vandersypen,~L.
  M.~K. {Spins in few-electron quantum dots}. \emph{Reviews of Modern Physics}
  \textbf{2007}, \emph{79}, 1217--1265\relax
\mciteBstWouldAddEndPuncttrue
\mciteSetBstMidEndSepPunct{\mcitedefaultmidpunct}
{\mcitedefaultendpunct}{\mcitedefaultseppunct}\relax
\EndOfBibitem
\bibitem[Lodari \latin{et~al.}(2019)Lodari, Tosato, Sabbagh, Schubert,
  Capellini, Sammak, Veldhorst, and Scappucci]{Lodari2019}
Lodari,~M.; Tosato,~A.; Sabbagh,~D.; Schubert,~M.~A.; Capellini,~G.;
  Sammak,~A.; Veldhorst,~M.; Scappucci,~G. {Light effective hole mass in
  undoped Ge/SiGe quantum wells}. \emph{Physical Review B} \textbf{2019},
  \emph{100}, 041304\relax
\mciteBstWouldAddEndPuncttrue
\mciteSetBstMidEndSepPunct{\mcitedefaultmidpunct}
{\mcitedefaultendpunct}{\mcitedefaultseppunct}\relax
\EndOfBibitem
\bibitem[Hendrickx \latin{et~al.}(2018)Hendrickx, Franke, Sammak, Kouwenhoven,
  Sabbagh, Yeoh, Li, Tagliaferri, Virgilio, Capellini, Scappucci, and
  Veldhorst]{Hendrickx2018}
Hendrickx,~N.~W.; Franke,~D.~P.; Sammak,~A.; Kouwenhoven,~M.; Sabbagh,~D.;
  Yeoh,~L.; Li,~R.; Tagliaferri,~M. L.~V.; Virgilio,~M.; Capellini,~G.;
  Scappucci,~G.; Veldhorst,~M. {Gate-controlled quantum dots and
  superconductivity in planar germanium}. \emph{Nature Communications}
  \textbf{2018}, \emph{9}, 2835\relax
\mciteBstWouldAddEndPuncttrue
\mciteSetBstMidEndSepPunct{\mcitedefaultmidpunct}
{\mcitedefaultendpunct}{\mcitedefaultseppunct}\relax
\EndOfBibitem
\bibitem[Hendrickx \latin{et~al.}(2019)Hendrickx, Franke, Sammak, Scappucci,
  and Veldhorst]{Hendrickx2019}
Hendrickx,~N.~W.; Franke,~D.~P.; Sammak,~A.; Scappucci,~G.; Veldhorst,~M. {Fast
  two-qubit logic with holes in germanium}. \emph{arXiv:1904.11443}
  \textbf{2019}, \relax
\mciteBstWouldAddEndPunctfalse
\mciteSetBstMidEndSepPunct{\mcitedefaultmidpunct}
{}{\mcitedefaultseppunct}\relax
\EndOfBibitem
\bibitem[Fischer \latin{et~al.}(2008)Fischer, Coish, Bulaev, and
  Loss]{Fischer2008}
Fischer,~J.; Coish,~W.~A.; Bulaev,~D.~V.; Loss,~D. {Spin decoherence of a heavy
  hole coupled to nuclear spins in a quantum dot}. \emph{Physical Review B}
  \textbf{2008}, \emph{78}, 155329\relax
\mciteBstWouldAddEndPuncttrue
\mciteSetBstMidEndSepPunct{\mcitedefaultmidpunct}
{\mcitedefaultendpunct}{\mcitedefaultseppunct}\relax
\EndOfBibitem
\bibitem[Testelin \latin{et~al.}(2009)Testelin, Bernardot, Eble, and
  Chamarro]{Testelin2009}
Testelin,~C.; Bernardot,~F.; Eble,~B.; Chamarro,~M. {Hole–spin dephasing time
  associated with hyperfine interaction in quantum dots}. \emph{Physical Review
  B} \textbf{2009}, \emph{79}, 195440\relax
\mciteBstWouldAddEndPuncttrue
\mciteSetBstMidEndSepPunct{\mcitedefaultmidpunct}
{\mcitedefaultendpunct}{\mcitedefaultseppunct}\relax
\EndOfBibitem
\bibitem[Petta(2005)]{Petta2005}
Petta,~J.~R. {Coherent Manipulation of Coupled Electron Spins in Semiconductor
  Quantum Dots}. \emph{Science} \textbf{2005}, \emph{309}, 2180--2184\relax
\mciteBstWouldAddEndPuncttrue
\mciteSetBstMidEndSepPunct{\mcitedefaultmidpunct}
{\mcitedefaultendpunct}{\mcitedefaultseppunct}\relax
\EndOfBibitem
\bibitem[Levy(2002)]{Levy2002}
Levy,~J. {Universal Quantum Computation with Spin- 1/2 Pairs and Heisenberg
  Exchange}. \emph{Physical Review Letters} \textbf{2002}, \emph{89},
  147902\relax
\mciteBstWouldAddEndPuncttrue
\mciteSetBstMidEndSepPunct{\mcitedefaultmidpunct}
{\mcitedefaultendpunct}{\mcitedefaultseppunct}\relax
\EndOfBibitem
\bibitem[Foletti \latin{et~al.}(2009)Foletti, Bluhm, Mahalu, Umansky, and
  Yacoby]{Foletti2009}
Foletti,~S.; Bluhm,~H.; Mahalu,~D.; Umansky,~V.; Yacoby,~A. {Universal quantum
  control of two-electron spin quantum bits using dynamic nuclear
  polarization}. \emph{Nature Physics} \textbf{2009}, \emph{5}, 903--908\relax
\mciteBstWouldAddEndPuncttrue
\mciteSetBstMidEndSepPunct{\mcitedefaultmidpunct}
{\mcitedefaultendpunct}{\mcitedefaultseppunct}\relax
\EndOfBibitem
\bibitem[Jock \latin{et~al.}(2018)Jock, Jacobson, Harvey-Collard, Mounce,
  Srinivasa, Ward, Anderson, Manginell, Wendt, Rudolph, Pluym, Gamble,
  Baczewski, Witzel, and Carroll]{Jock2018}
Jock,~R.~M.; Jacobson,~N.~T.; Harvey-Collard,~P.; Mounce,~A.~M.; Srinivasa,~V.;
  Ward,~D.~R.; Anderson,~J.; Manginell,~R.; Wendt,~J.~R.; Rudolph,~M.;
  Pluym,~T.; Gamble,~J.~K.; Baczewski,~A.~D.; Witzel,~W.~M.; Carroll,~M.~S. {A
  silicon metal-oxide-semiconductor electron spin-orbit qubit}. \emph{Nature
  Communications} \textbf{2018}, \emph{9}, 1768\relax
\mciteBstWouldAddEndPuncttrue
\mciteSetBstMidEndSepPunct{\mcitedefaultmidpunct}
{\mcitedefaultendpunct}{\mcitedefaultseppunct}\relax
\EndOfBibitem
\bibitem[Terrazos \latin{et~al.}(2018)Terrazos, Marcellina, Coppersmith,
  Friesen, Hamilton, Hu, Koiller, Saraiva, Culcer, and Capaz]{Terrazos2018}
Terrazos,~L.~A.; Marcellina,~E.; Coppersmith,~S.~N.; Friesen,~M.;
  Hamilton,~A.~R.; Hu,~X.; Koiller,~B.; Saraiva,~A.~L.; Culcer,~D.;
  Capaz,~R.~B. {Qubits Based on Hole Quantum Dots in Strained Ge}.
  \textbf{2018}, \relax
\mciteBstWouldAddEndPunctfalse
\mciteSetBstMidEndSepPunct{\mcitedefaultmidpunct}
{}{\mcitedefaultseppunct}\relax
\EndOfBibitem
\bibitem[Morrison \latin{et~al.}(2014)Morrison, Wi{\'{s}}niewski, Rhead,
  Foronda, Leadley, and Myronov]{Morrison2014}
Morrison,~C.; Wi{\'{s}}niewski,~P.; Rhead,~S.~D.; Foronda,~J.; Leadley,~D.~R.;
  Myronov,~M. {Observation of Rashba zero-field spin splitting in a strained
  germanium 2D hole gas}. \emph{Applied Physics Letters} \textbf{2014},
  \emph{105}, 182401\relax
\mciteBstWouldAddEndPuncttrue
\mciteSetBstMidEndSepPunct{\mcitedefaultmidpunct}
{\mcitedefaultendpunct}{\mcitedefaultseppunct}\relax
\EndOfBibitem
\bibitem[Hardy \latin{et~al.}(2019)Hardy, Harris, Su, Chuang, Moussa, Maurer,
  Li, Lu, and Luhman]{Hardy2019}
Hardy,~W.~J.; Harris,~C.~T.; Su,~Y.-H.; Chuang,~Y.; Moussa,~J.; Maurer,~L.~N.;
  Li,~J.-Y.; Lu,~T.-M.; Luhman,~D.~R. {Single and double hole quantum dots in
  strained Ge/SiGe quantum wells}. \emph{Nanotechnology} \textbf{2019},
  \emph{30}, 215202\relax
\mciteBstWouldAddEndPuncttrue
\mciteSetBstMidEndSepPunct{\mcitedefaultmidpunct}
{\mcitedefaultendpunct}{\mcitedefaultseppunct}\relax
\EndOfBibitem
\bibitem[R{\"{o}}ssner \latin{et~al.}(2004)R{\"{o}}ssner, Chrastina, Isella,
  and von K{\"{a}}nel]{Rossner2004}
R{\"{o}}ssner,~B.; Chrastina,~D.; Isella,~G.; von K{\"{a}}nel,~H. {Scattering
  mechanisms in high-mobility strained Ge channels}. \emph{Applied Physics
  Letters} \textbf{2004}, \emph{84}, 3058--3060\relax
\mciteBstWouldAddEndPuncttrue
\mciteSetBstMidEndSepPunct{\mcitedefaultmidpunct}
{\mcitedefaultendpunct}{\mcitedefaultseppunct}\relax
\EndOfBibitem
\bibitem[Ihn(2009)]{Ihn2009}
Ihn,~T. \emph{{Semiconductor Nanostructures}}; Oxford University Press,
  2009\relax
\mciteBstWouldAddEndPuncttrue
\mciteSetBstMidEndSepPunct{\mcitedefaultmidpunct}
{\mcitedefaultendpunct}{\mcitedefaultseppunct}\relax
\EndOfBibitem
\bibitem[NSM@mail.ioffe.ru()]{NSM_IOFFE}
NSM@mail.ioffe.ru, {Basic Parameters of Germanium (Ge)}.
  \url{http://www.ioffe.ru/SVA/NSM/Semicond/Ge/basic.html}\relax
\mciteBstWouldAddEndPuncttrue
\mciteSetBstMidEndSepPunct{\mcitedefaultmidpunct}
{\mcitedefaultendpunct}{\mcitedefaultseppunct}\relax
\EndOfBibitem
\bibitem[Livermore \latin{et~al.}(1996)Livermore, Crouch, Westervelt, Campman,
  and Gossard]{Livermore1996}
Livermore,~C.; Crouch,~C.~H.; Westervelt,~R.~M.; Campman,~K.~L.; Gossard,~A.~C.
  {The Coulomb Blockade in Coupled Quantum Dots}. \emph{Science (New York,
  N.Y.)} \textbf{1996}, \emph{274}, 1332--5\relax
\mciteBstWouldAddEndPuncttrue
\mciteSetBstMidEndSepPunct{\mcitedefaultmidpunct}
{\mcitedefaultendpunct}{\mcitedefaultseppunct}\relax
\EndOfBibitem
\bibitem[Matveev \latin{et~al.}(1996)Matveev, Glazman, and
  Baranger]{Matveev1996}
Matveev,~K.~A.; Glazman,~L.~I.; Baranger,~H.~U. {Coulomb blockade of tunneling
  through a double quantum dot}. \emph{Physical Review B} \textbf{1996},
  \emph{54}, 5637--5646\relax
\mciteBstWouldAddEndPuncttrue
\mciteSetBstMidEndSepPunct{\mcitedefaultmidpunct}
{\mcitedefaultendpunct}{\mcitedefaultseppunct}\relax
\EndOfBibitem
\bibitem[Ono \latin{et~al.}(2002)Ono, Austing, Tokura, and Tarucha]{Ono2002}
Ono,~K.; Austing,~D.~G.; Tokura,~Y.; Tarucha,~S. {Current rectification by
  Pauli exclusion in a weakly coupled double quantum dot system.} \emph{Science
  (New York, N.Y.)} \textbf{2002}, \emph{297}, 1313--7\relax
\mciteBstWouldAddEndPuncttrue
\mciteSetBstMidEndSepPunct{\mcitedefaultmidpunct}
{\mcitedefaultendpunct}{\mcitedefaultseppunct}\relax
\EndOfBibitem
\bibitem[Field \latin{et~al.}(1993)Field, Smith, Pepper, Ritchie, Frost, Jones,
  and Hasko]{Field1993}
Field,~M.; Smith,~C.~G.; Pepper,~M.; Ritchie,~D.~A.; Frost,~J. E.~F.; Jones,~G.
  A.~C.; Hasko,~D.~G. {Measurements of Coulomb blockade with a noninvasive
  voltage probe}. \emph{Physical Review Letters} \textbf{1993}, \emph{70},
  1311--1314\relax
\mciteBstWouldAddEndPuncttrue
\mciteSetBstMidEndSepPunct{\mcitedefaultmidpunct}
{\mcitedefaultendpunct}{\mcitedefaultseppunct}\relax
\EndOfBibitem
\bibitem[Franceschi and van~der Wiel(2010)Franceschi, and van~der
  Wiel]{Franceschi2010}
Franceschi,~S.~D.; van~der Wiel,~W.~G. In \emph{Handbook of Nanophysics:
  Nanoparticles and Quantum Dots}; Sattler,~K.~D., Ed.; CRC Press, 2010;
  Chapter Kondo Effe, pp 646--664\relax
\mciteBstWouldAddEndPuncttrue
\mciteSetBstMidEndSepPunct{\mcitedefaultmidpunct}
{\mcitedefaultendpunct}{\mcitedefaultseppunct}\relax
\EndOfBibitem
\bibitem[Li \latin{et~al.}(2015)Li, Hudson, Dzurak, and Hamilton]{Li2015}
Li,~R.; Hudson,~F.~E.; Dzurak,~A.~S.; Hamilton,~A.~R. {Pauli Spin Blockade of
  Heavy Holes in a Silicon Double Quantum Dot}. \emph{Nano Letters}
  \textbf{2015}, \emph{15}, 7314--7318\relax
\mciteBstWouldAddEndPuncttrue
\mciteSetBstMidEndSepPunct{\mcitedefaultmidpunct}
{\mcitedefaultendpunct}{\mcitedefaultseppunct}\relax
\EndOfBibitem
\bibitem[Csonka \latin{et~al.}(2008)Csonka, Hofstetter, Freitag, Oberholzer,
  Sch{\"{o}}nenberger, Jespersen, Aagesen, and Nyg{\aa}rd]{Csonka2008}
Csonka,~S.; Hofstetter,~L.; Freitag,~F.; Oberholzer,~S.;
  Sch{\"{o}}nenberger,~C.; Jespersen,~T.~S.; Aagesen,~M.; Nyg{\aa}rd,~J. {Giant
  Fluctuations and Gate Control of the g-Factor in InAs Nanowire Quantum Dots}.
  \emph{Nano Letters} \textbf{2008}, \emph{8}, 3932--3935\relax
\mciteBstWouldAddEndPuncttrue
\mciteSetBstMidEndSepPunct{\mcitedefaultmidpunct}
{\mcitedefaultendpunct}{\mcitedefaultseppunct}\relax
\EndOfBibitem
\bibitem[Petta and Ralph(2002)Petta, and Ralph]{Petta2002}
Petta,~J.~R.; Ralph,~D.~C. {Measurements of Strongly Anisotropic g Factors for
  Spins in Single Quantum States}. \emph{Physical Review Letters}
  \textbf{2002}, \emph{89}, 156802\relax
\mciteBstWouldAddEndPuncttrue
\mciteSetBstMidEndSepPunct{\mcitedefaultmidpunct}
{\mcitedefaultendpunct}{\mcitedefaultseppunct}\relax
\EndOfBibitem
\bibitem[Brouwer \latin{et~al.}(2000)Brouwer, Waintal, and
  Halperin]{Brouwer2000}
Brouwer,~P.~W.; Waintal,~X.; Halperin,~B.~I. {Fluctuating Spin g -Tensor in
  Small Metal Grains}. \emph{Physical Review Letters} \textbf{2000}, \emph{85},
  369--372\relax
\mciteBstWouldAddEndPuncttrue
\mciteSetBstMidEndSepPunct{\mcitedefaultmidpunct}
{\mcitedefaultendpunct}{\mcitedefaultseppunct}\relax
\EndOfBibitem
\bibitem[van Kesteren \latin{et~al.}(1990)van Kesteren, Cosman, van~der Poel,
  and Foxon]{VanKesteren1990}
van Kesteren,~H.~W.; Cosman,~E.~C.; van~der Poel,~W. A. J.~A.; Foxon,~C.~T.
  {Fine structure of excitons in type-II GaAs/AlAs quantum wells}.
  \emph{Physical Review B} \textbf{1990}, \emph{41}, 5283--5292\relax
\mciteBstWouldAddEndPuncttrue
\mciteSetBstMidEndSepPunct{\mcitedefaultmidpunct}
{\mcitedefaultendpunct}{\mcitedefaultseppunct}\relax
\EndOfBibitem
\bibitem[Voisin \latin{et~al.}(2016)Voisin, Maurand, Barraud, Vinet, Jehl,
  Sanquer, Renard, and {De Franceschi}]{Voisin2016}
Voisin,~B.; Maurand,~R.; Barraud,~S.; Vinet,~M.; Jehl,~X.; Sanquer,~M.;
  Renard,~J.; {De Franceschi},~S. {Electrical Control of g -Factor in a
  Few-Hole Silicon Nanowire MOSFET}. \emph{Nano Letters} \textbf{2016},
  \emph{16}, 88--92\relax
\mciteBstWouldAddEndPuncttrue
\mciteSetBstMidEndSepPunct{\mcitedefaultmidpunct}
{\mcitedefaultendpunct}{\mcitedefaultseppunct}\relax
\EndOfBibitem
\bibitem[Vigneau \latin{et~al.}(2019)Vigneau, Mizokuchi, Zanuz, Huang, Tan,
  Maurand, Frolov, Sammak, Scappucci, Lefloch, and {De
  Franceschi}]{Vigneau2019}
Vigneau,~F.; Mizokuchi,~R.; Zanuz,~D.~C.; Huang,~X.; Tan,~S.; Maurand,~R.;
  Frolov,~S.; Sammak,~A.; Scappucci,~G.; Lefloch,~F.; {De Franceschi},~S.
  {Germanium Quantum-Well Josephson Field-Effect Transistors and
  Interferometers}. \emph{Nano Letters} \textbf{2019}, \emph{19},
  1023--1027\relax
\mciteBstWouldAddEndPuncttrue
\mciteSetBstMidEndSepPunct{\mcitedefaultmidpunct}
{\mcitedefaultendpunct}{\mcitedefaultseppunct}\relax
\EndOfBibitem
\bibitem[Hendrickx \latin{et~al.}(2019)Hendrickx, Tagliaferri, Kouwenhoven, Li,
  Franke, Sammak, Brinkman, Scappucci, and Veldhorst]{Hendrickx2019a}
Hendrickx,~N.~W.; Tagliaferri,~M. L.~V.; Kouwenhoven,~M.; Li,~R.;
  Franke,~D.~P.; Sammak,~A.; Brinkman,~A.; Scappucci,~G.; Veldhorst,~M.
  {Ballistic supercurrent discretization and micrometer-long Josephson coupling
  in germanium}. \emph{Physical Review B} \textbf{2019}, \emph{99},
  075435\relax
\mciteBstWouldAddEndPuncttrue
\mciteSetBstMidEndSepPunct{\mcitedefaultmidpunct}
{\mcitedefaultendpunct}{\mcitedefaultseppunct}\relax
\EndOfBibitem
\bibitem[Mizokuchi \latin{et~al.}(2018)Mizokuchi, Maurand, Vigneau, Myronov,
  and {De Franceschi}]{Mizokuchi2018}
Mizokuchi,~R.; Maurand,~R.; Vigneau,~F.; Myronov,~M.; {De Franceschi},~S.
  {Ballistic One-Dimensional Holes with Strong g-Factor Anisotropy in
  Germanium}. \emph{Nano Letters} \textbf{2018}, \emph{18}, 4861--4865\relax
\mciteBstWouldAddEndPuncttrue
\mciteSetBstMidEndSepPunct{\mcitedefaultmidpunct}
{\mcitedefaultendpunct}{\mcitedefaultseppunct}\relax
\EndOfBibitem
\bibitem[Lutchyn \latin{et~al.}(2010)Lutchyn, Sau, and {Das
  Sarma}]{Lutchyn2010}
Lutchyn,~R.~M.; Sau,~J.~D.; {Das Sarma},~S. {Majorana Fermions and a
  Topological Phase Transition in Semiconductor-Superconductor
  Heterostructures}. \emph{Physical Review Letters} \textbf{2010}, \emph{105},
  077001\relax
\mciteBstWouldAddEndPuncttrue
\mciteSetBstMidEndSepPunct{\mcitedefaultmidpunct}
{\mcitedefaultendpunct}{\mcitedefaultseppunct}\relax
\EndOfBibitem
\bibitem[Oreg \latin{et~al.}(2010)Oreg, Refael, and von Oppen]{Oreg2010}
Oreg,~Y.; Refael,~G.; von Oppen,~F. {Helical Liquids and Majorana Bound States
  in Quantum Wires}. \emph{Physical Review Letters} \textbf{2010}, \emph{105},
  177002\relax
\mciteBstWouldAddEndPuncttrue
\mciteSetBstMidEndSepPunct{\mcitedefaultmidpunct}
{\mcitedefaultendpunct}{\mcitedefaultseppunct}\relax
\EndOfBibitem
\bibitem[Lawrie \latin{et~al.}(2019)Lawrie, Eenink, Hendrickx, Boter, Petit,
  Amitonov, Lodari, Wuetz, Volk, Philips, Droulers, Kalhor, van Riggelen,
  Brousse, Sammak, Vandersypen, Scappucci, and Veldhorst]{Lawrie2019}
Lawrie,~W. I.~L. \latin{et~al.}  {Quantum Dot Arrays in Silicon and Germanium}.
  \emph{arXiv:1909.06575} \textbf{2019}, \relax
\mciteBstWouldAddEndPunctfalse
\mciteSetBstMidEndSepPunct{\mcitedefaultmidpunct}
{}{\mcitedefaultseppunct}\relax
\EndOfBibitem
\end{mcitethebibliography}

\end{document}